\documentclass[preprint,aps]{revtex4}
\usepackage{mathrsfs}

\usepackage{graphicx}
\usepackage{multirow}
\usepackage{makecell}
\usepackage{booktabs}
\begin{document}

\title{Genuine Electronic Structure and Superconducting Gap Structure in (Ba$_{0.6}$K$_{0.4}$)Fe$_2$As$_2$ Superconductor}

\author{Yongqing Cai$^{1,2,\sharp}$, Jianwei Huang$^{1,\sharp}$, Taimin Miao$^{1,2}$, Dingsong Wu$^{1,2}$, Qiang Gao$^{1,2}$, Cong Li$^{1,2}$, Yu Xu$^{1,2}$, Junjie Jia$^{1,2}$, Qingyan Wang$^{1,2,3}$, Yuan Huang$^{1,2,3}$, Guodong Liu$^{1,2,3}$, Fengfeng Zhang$^{4}$, Shenjin Zhang$^{4}$, Feng Yang$^{4}$, Zhimin Wang$^{4}$, Qinjun Peng$^{4}$, Zuyan Xu$^{4}$, Lin Zhao$^{1,2,3*}$ and X. J. Zhou$^{1,2,3,5,*}$}


\affiliation{
\\$^{1}$National Lab for Superconductivity, Beijing National Laboratory for Condensed Matter Physics, Institute of Physics,
Chinese Academy of Sciences, Beijing 100190, China
\\$^{2}$University of Chinese Academy of Sciences, Beijing 100049, China
\\$^{3}$ Songshan Lake Materials Laboratory, Dongguan 523808, China
\\$^{4}$Technical Institute of Physics and Chemistry, Chinese Academy of Sciences, Beijing 100190, China
\\$^{5}$Beijing Academy of Quantum Information Sciences, Beijing 100193, China
\\$^{\sharp}$These authors contribute equally to the present work.
\\$^{*}$Corresponding author: lzhao@iphy.ac.cn and XJZhou@iphy.ac.cn
}

\date{March 24, 2021}

\maketitle

{\bf
The electronic structure and superconducting gap structure are prerequisites to establish microscopic theories in understanding the superconductivity mechanism of iron-based superconductors. However, even for the most extensively studied optimally-doped (Ba$_{0.6}$K$_{0.4}$)Fe$_2$As$_2$, there remain outstanding controversies on its electronic structure and superconducting gap structure. Here we resolve these issues by carrying out high-resolution angle-resolved photoemission spectroscopy (ARPES) measurements on the optimally-doped (Ba$_{0.6}$K$_{0.4}$)Fe$_2$As$_2$ superconductor using both Helium lamp and laser light sources. Our results indicate the "flat band" feature observed around the Brillouin zone center in the superconducting state originates from the combined effect of the superconductivity-induced band back-bending and the folding of a band from the zone corner to the center. We found direct evidence of the band folding between the zone corner and the center in both the normal and superconducting state. Our resolution of the origin of the flat band makes it possible to assign the three hole-like bands around the zone center and determine their superconducting gap correctly. Around the zone corner, we observe a tiny electron-like band and an M-shaped band simultaneously in both the normal and superconducting states. The obtained gap size for the bands around the zone corner ($\sim$5.5\,meV) is significantly smaller than all the previous ARPES measurements. Our results establish a new superconducting gap structure around the zone corner and resolve a number of prominent controversies concerning the electronic structure and superconducting gap structure in the optimally-doped (Ba$_{0.6}$K$_{0.4}$)Fe$_2$As$_2$. They provide new insights in examining and establishing theories in understanding superconductivity mechanism in iron-based superconductors.
}

\vspace{3mm}

Since the discovery of superconductivity in iron-based superconductors in 2008
\cite{HHosono2008YKamihara,ZXZhao2008ZARen,MKWu2008FCHsu,DJohrendt2008MRotter,CQJin2008XCWang},
the superconductivity mechanism remains under hot debate in spite of great experimental and theoretical efforts
\cite{HHosono2009KIshida, RLGreene2010JPaglione, IIMazin2011Hirschfeld, AVChubukov2017RMFernandes}.
Understanding the electronic structure and superconducting gap structure is the prerequisite to establish microscopic theories to understand the superconductivity mechanism of the iron-based superconductors. Angle-resolved photoemission spectroscopy (ARPES) has provided key information in studying the iron-based superconductors\cite{HDing2011PRichard,DLFeng2013ZRYe,XJZhou2015XLiu,XJZhou2018LZhao,ZXShen2020JASobota,XJZhou2021LZhao}.
However, even for the prototypical (Ba,K)Fe$_2$As$_2$ superconductor that has been most extensively studied by ARPES\cite{XJZhou2008LZhao, HDing2008NLWang, HDing2009PRichard, MZHasan2008LWray, SVBorisenko2009DVEvtushinsky, SVBorisenko2009AKoitzsch, SVBorisenko2009VBZabolotnyyNature, DLFeng2010YZhang, HDing2011YMXu, NLWang2011HDing, SShin2011TShimojima, SVBorisenko2011DVEvtushinsky, SShin2012WMalaeb, KIshizaka2012TShimojima, SVBorisenko2014DVEvtushinsky, SShin2014WMalaeb, HDing2014PZhang, SShin2017TShimojima, XJZhou2019JWhuang}, there remain outstanding controversies on its electronic structure and superconducting gap structure.
First, regarding the electronic structure around the $\Gamma$ point, a prominent issue concerns the origin of the "flat band" in the superconducting state. The ARPES measurements revealed that a flat band appears near the $\Gamma$ point in the superconducting state of the optimally-doped (Ba$_{0.6}$K$_{0.4}$)Fe$_2$As$_2$ (Fig. 1b)\cite{XJZhou2008LZhao, DLFeng2010YZhang, SVBorisenko2011DVEvtushinsky, HDing2014PZhang, SShin2017TShimojima}. This flat band has been attributed to the formation of Bogoliubons\cite{SVBorisenko2011DVEvtushinsky}(Fig. 1d), appearance of the in-gap state\cite{HDing2014PZhang}(Fig. 1e), the superconductivity-induced back-bending of a band\cite{DLFeng2010YZhang}(Fig. 1f), or the folding of the band at M to $\Gamma$\cite{SShin2017TShimojima}(Fig. 1g). The resolution of this controversy is directly related to the assignment of the three hole-like bands around $\Gamma$ and their superconducting gap determination.
Second, there has been a long-time debate on the Fermi surface topology and electronic structure around the M point for the optimally-doped (Ba$_{0.6}$K$_{0.4}$)Fe$_2$As$_2$. It has been controversial on whether the Fermi surface around the M point consists of only a tiny electron pocket\cite{XJZhou2008LZhao}, two sizable electron pockets\cite{NLWang2011HDing}, or propeller-shaped Fermi surface with one tiny electron pocket at M and four hole-like pockets around M\cite{SVBorisenko2009VBZabolotnyyNature, SVBorisenko2009DVEvtushinsky}.
Third, there are contradicting results on the superconducting gap structure even for the same optimally-doped (Ba$_{0.6}$K$_{0.4}$)Fe$_2$As$_2$. For the three hole-like bands around the $\Gamma$ point, most ARPES measurements report Fermi surface sheet-dependent superconducting gap\cite{XJZhou2008LZhao, HDing2008NLWang, MZHasan2008LWray, SVBorisenko2009DVEvtushinsky, DLFeng2010YZhang, HDing2011YMXu, XJZhou2019JWhuang} while some ARPES measurements found that the superconducting gap is independent of the Fermi surface sheets with much smaller magnitude\cite{SShin2011TShimojima, SShin2012WMalaeb}. Regarding the superconducting gap structure around the M point, because of the controversy of the Fermi surface topology, it remains unsettled\cite{XJZhou2008LZhao, HDing2008NLWang, MZHasan2008LWray, SVBorisenko2009DVEvtushinsky, TTakahashi2009KNakayama, SShin2017TShimojima}.

In this paper, we carried out high-resolution ARPES measurements on the optimally-doped (Ba$_{0.6}$K$_{0.4}$)Fe$_2$As$_2$ to resolve the detailed electronic structure and superconducting gap structure by using both Helium lamp and laser light sources. We observe a clear "flat band" feature around the $\Gamma$ point in the superconducting state and reveal its origin. We found direct evidence of the band folding between M and $\Gamma$. Near the $\Gamma$ point, our resolution of the origin of the flat band makes it possible to assign the three hole-like bands and determine their superconducting gap correctly. Near the M point, we observe both a tiny electron-like band $\delta$ and an M-shaped $\varepsilon$ band simultaneously in the normal and superconducting states. This makes it possible to establish a new superconducting gap structure around the M point. Our results resolve a number of significant controversies concerning the genuine electronic structure and superconducting gap structure in the prototypical iron-based superconductor, the optimally-doped (Ba$_{0.6}$K$_{0.4}$)Fe$_2$As$_2$.

The optimally-doped (Ba$_{0.6}$K$_{0.4}$)Fe$_2$As$_2$ single crystals were grown using a FeAs flux method with excess FeAs as flux \cite{GFChen2008JLLuo}. It shows a superconducting transition temperature T$_C$ of $\sim$ 38 K with a narrow transition width of 0.5 K from the magnetic susceptibility measurement\cite{XJZhou2019JWhuang}. High resolution angle-resolved photoemission measurements were carried out on our two lab-based ARPES systems\cite{XJZhou2008GDLiu,XJZhou2018}. One is equipped with Helium discharge lamp as the light source which can provide a photon energy of \emph{h}$\nu$ = 21.21818 eV (Helium I), and a Scienta DA30L electron energy analyzer. It can cover a large momentum space so that we can get the overall Fermi surface and band structures at both the Brillouin zone center and the zone corners. The energy resolution was set at 10 meV for the Fermi surface mapping and at 4 meV for the band structure measurements, and the angular resolution is ~$\sim$ 0.3 $^\circ$. The other system is equipped with the 6.994\,eV vacuum-ultra-violet (VUV) laser and angle-resolved time-of-flight electron energy analyzer (ARToF 10k by Scienta Omicron). It can cover two-dimensional momentum space simultaneously and has much weaker non-linearity effect so that the intrinsic signal can be measured. The energy resolution was set at 1 meV and the angular resolution was 0.3 $^\circ$ corresponding to 0.004 ${\AA}^{-1}$ momentum resolution at the photon energy of 6.994\,eV. The detailed Fermi surface and band structure around the $\Gamma$ point were measured in this system. All the samples were cleaved \emph{in situ} at 18\,K and measured in ultrahigh vacuum with a base pressure better than 5$\times$10$^{-11}$ mbar. The Fermi level is referenced by measuring on a clean polycrystalline gold that is electrically connected to the sample, as well as the normal state measurement of the sample. We have performed measurements on multiple samples and the results are reproducible.

Figure 1a shows the overall Fermi surface mapping of the optimally-doped (Ba$_{0.6}$K$_{0.4}$)Fe$_2$As$_2$ superconductor measured at 18\,K in the superconducting state using Helium discharge lamp. Two hole-like Fermi pockets around $\Gamma$ and strong spots around M are observed, which is consistent with the previous results\cite{XJZhou2008LZhao,MZHasan2008LWray,SVBorisenko2009VBZabolotnyyNature}. Fig. 1b shows the band structure measured at 18\,K along the $\Gamma$-M$_1$ direction crossing the $\Gamma$ point (Cut 1 in Fig. 1a). In addition to the two hole-like bands, the inner and outer ones, a prominent feature is the "flat band" observed around the $\Gamma$ point, as marked by the dashed red rectangular frame in Fig. 1b. Fig. 1c shows the corresponding photoemission spectra (energy distribution curves, EDCs) where the position of the flat band is marked and plotted in Fig. 1b. It is found that this flat band can cross the inner hole-like band and extend to the outer hole-like band. Such a flat band has been commonly observed in the previous measurements\cite{XJZhou2008LZhao, DLFeng2010YZhang, SVBorisenko2011DVEvtushinsky, HDing2014PZhang, SShin2017TShimojima}. However, the origin of this flat band remains under debate and a number of possible scenarios have been proposed. First, the abnormal strong intensity of the flat band near the $\Gamma$ point is attributed to the fusion of Bogoliubov dispersions (Bogoliubons) which is due to comparability of all relevant energy scales—electronic band energy, pairing energy, and energy of a mode, as schematically shown in Fig. 1d\cite{SVBorisenko2011DVEvtushinsky}. The second possibility is associated with the formation of "in-gap" state inside the superconducting gap likely due to disorder scattering\cite{HDing2014PZhang}. In this case, the flat band is considered to be due to the Bogoliubov back-bending of the inner hole-like band (Fig. 1e)\cite{HDing2014PZhang}. The third possibility is shown in Fig. 1f. The inner hole-like band consists of two degenerate bands ($\alpha$ and $\beta$) in the normal state; they exhibit different superconducting gap size in the superconducting state\cite{DLFeng2010YZhang}. The Bogoliubov back-bending from the $\beta$ band, which shows a big superconducting gap, forms the spectral enhancement and flat band. The fourth possibility is the folded electronic structure from the M point due to suggestion of an antiferroic electronic instability, coexisting with the superconductivity in the nonmagnetic state of (Ba$_{0.6}$K$_{0.4}$)Fe$_2$As$_2$ superconductor\cite{SShin2017TShimojima}. The flat band crossing with the $\alpha$ band is from folding of the flat band around the M point with a folding vector (0,0)-($\pm\pi$,$\pm\pi$), as shown in Fig. 1g. Above four possibilities all can produce the spectral weight enhancement and flat band features, but the underlying physics is different. The clarification on the origin of the flat band near the $\Gamma$ point is important not only for arriving at the correct electronic structure and related superconducting gap determination, but also for understanding the underlying physics in (Ba$_{0.6}$K$_{0.4}$)Fe$_2$As$_2$ superconductor.

In order to understand the origin of the flat band, we also carried out high resolution laser-based ARPES measurements on the optimally-doped (Ba$_{0.6}$K$_{0.4}$)Fe$_2$As$_2$ superconductor. Fig. 2a shows the Fermi surface mapping measured on the fresh sample at 13\,K in the superconducting state. For comparison, Fig. 2b shows the Fermi surface mapping on the aged sample measured under the same condition. In both cases, the two hole-like Fermi surface sheets are clearly observed. Figs. 2c-2d show the band structure measured along the same $\Gamma$-M$_1$ direction crossing the $\Gamma$ point for the fresh and aged samples. The two hole-like bands are clearly observed although their intensity is suppressed in the aged sample (Fig. 2d). In addition, a flat band is present in both the fresh and aged samples at $\sim$15 meV and spans a wide momentum space between the $\Gamma$ point and the outer hole-like band crossing the inner hole-like band. Such a flat band can be more clearly seen in the corresponding second derivative images shown in Figs. 2e-2f. Figs. 2g-2h show photoemission spectra (EDCs) for the band structure of the fresh sample in Fig. 2c while the EDCs for the aged sample (Fig. 2d) are shown in Fig. 2i. For the fresh sample, the EDCs at the Fermi momenta (k$_{\beta}$ and k$_{\gamma}$ in Figs. 2g-2h) exhibit sharp superconductivity-induced coherence peaks. Such coherence peaks are strongly suppressed in the aged sample (Fig. 2i). In Figs. 2g-2i, the position of the flat band in the EDCs is marked by the red tips and the obtained dispersion of the flat band in the fresh and aged samples is plotted in Fig. 2e and Fig. 2f, respectively. Figs. 2j-2m compare directly the EDCs for the fresh and aged samples at several typical momenta. While the sharp coherence peaks in the fresh sample become strongly suppressed in the aged sample (Fig. 2k and Fig. 2m), the flat band is quite similar in the fresh and aged samples both in its intensity and energy position (Figs. 2j-2m). In the aged sample, the intensity of the main bands and the flat band becomes comparable (Fig. 2k and Fig. 2m), facilitating the identification of the flat band and understanding its origin. Our results indicate that the sample aging mainly affects the superconducting coherence peaks. We note that some previous laser ARPES measurement results of the optimally-doped (Ba$_{0.6}$K$_{0.4}$)Fe$_2$As$_2$ superconductor\cite{SShin2011TShimojima,SShin2017TShimojima} are similar to the results we obtained on the aged samples.

In both our Helium lamp (Fig. 1) and laser (Fig. 2) ARPES measurements, we find that the flat band around the $\Gamma$ point spans a wide momentum space crossing the inner hole-like band. These clearly indicate that the Bogoliubov back-bending of the inner hole-like band alone can not account for the formation of such a flat band, as proposed in Fig. 1d, Fig. 1e and Fig. 1f. In our Helium lamp ARPES measurements in Fig. 1b, we observe clear superconductivity-induced coherence peak (Fig. 1c). In this case, we do not observe the "in-gap" state feature for the inner hole-like band as reported before\cite{HDing2014PZhang}. In our laser ARPES measurements on the fresh sample in Fig. 2c, when the inner hole-like band is dominant and sharp superconducting coherence peaks are observed (Figs. 2g-2h), we do not observe any signature of the "in-gap" state either. In our laser ARPES measurements on the aged sample in Fig. 2d, when the intensity of the inner hole-like band and flat band is comparable (Figs. 2j-2m), the observed band looks similar to that reported before where a part of the band near the Fermi level seems to be separated from the rest of the inner hole-like band\cite{HDing2014PZhang}. Careful examination of the band in Fig. 2d and its comparison with that in Fig. 2c indicates that, even in this case, there is no formation of the "in-gap" state. First, because the flat band crosses the inner band (Fig. 2d), the flat band can not be produced from the back-bending of the inner $\beta$ band. Second, the crossing of the inner band and the flat band makes the overlapping area stronger (Fig. 2d); this produces spectral dips in the EDCs around the Fermi momenta of the inner band (Fig. 2i), making the part of the inner band appear isolated from the rest of the band.

In order to resolve the controversies on the electronic structure\cite{XJZhou2008LZhao,NLWang2011HDing,SVBorisenko2009VBZabolotnyyNature, SVBorisenko2009DVEvtushinsky} and the associated superconducting gap structure\cite{XJZhou2008LZhao, HDing2008NLWang, MZHasan2008LWray, SVBorisenko2009DVEvtushinsky, DLFeng2010YZhang, HDing2011YMXu, XJZhou2019JWhuang} in the optimally-doped (Ba$_{0.6}$K$_{0.4}$)Fe$_2$As$_2$ superconductor, in particular to examine possible band folding between $\Gamma$ and M\cite{SShin2017TShimojima}, we carried out detailed ARPES measurements around the M point in both the normal and superconducting states. Fig. 3 shows the band structure measured crossing the M$_2$ point along two high-symmetry $\Gamma$-M$_1$ (Figs. 3a-3c at 45\,K in the normal state and Figs. 3d-3f at 18\,K in the superconducting state) and $\Gamma$-X (Figs. 3g-3i at 45\,K in the normal state and Figs. 3j-3l at 18\,K in the superconducting state) directions. In the measured band structures along the $\Gamma$-M$_1$ direction (Figs. 3a-3f), two bands are observed. One is a tiny electron-like band (named $\delta$ in Fig. 3) which lies with its bottom just touching the Fermi level in the normal state (marked by the cyan curve in Fig. 3c). It forms a short section of flat band in the superconducting state (marked by the cyan curve in Fig. 3f). The other band is the M-shaped $\varepsilon$ band as marked by the red curves in Fig. 3c and Fig. 3f. In the measured band structures along the $\Gamma$-X direction (Figs. 3g-3l), these two coexisting bands are also observed. The tiny electron-like $\delta$ band in the normal state (marked by the cyan curve in Fig. 3i) turns into a section of flat band in the superconducting state (marked by the cyan curve in Fig. 3l). But in this case, the $\varepsilon$ band shows up as a section of electron-like band in both the normal and superconducting states, as marked by the red curves in Fig. 3i and Fig. 3l. The strong spots around the M point in the Fermi surface mapping (Fig. 1a), which are located along the $\Gamma$-M directions, can be understood because they originate from the M-shaped $\varepsilon$ band which appears only along the $\Gamma$-M directions (Fig. 3c and Fig. 3f), not along the $\Gamma$-X direction (Fig. 3i and Fig. 3l).

Figure 3m and Fig. 3n show photoemission spectra (EDCs) measured in the normal and superconducting states at several typical momentum points along the $\Gamma$-M$_1$ and $\Gamma$-X directions, respectively. It is found that the normal state of the optimally-doped (Ba$_{0.6}$K$_{0.4}$)Fe$_2$As$_2$ superconductor is quite incoherent. Except for a small peak in the EDCs at the M point, no quasiparticle peaks are observed in the other EDCs in the normal state. This is consistent with the previous observation of incoherent state around the $\Gamma$ point in the normal state\cite{XJZhou2019JWhuang}. Upon entering the superconducting state, both the $\delta$ band and the $\varepsilon$ band form sharp superconducting coherence peaks (Fig. 3m and Fig. 3n). In particular, sharp coherence peaks form along the M-shaped $\varepsilon$ band (Fig. 3c) over a wide momentum space (Fig. 3m). The electronic states around the M point are dramatically driven to transform from the incoherent normal state into a highly coherent superconducting state.

The simultaneous observation of both the $\delta$ band and the $\varepsilon$ band, in both the normal and superconducting states, makes it possible to make a proper band assignment and a precise superconducting gap determination. Fig. 3o and Fig. 3p show the symmetrized EDCs at the M$_2$ point in the normal (black curves) and superconducting (red curves) states from the bands measured along the $\Gamma$-M$_1$ (Fig. 3a and Fig. 3d) and $\Gamma$-X (Fig. 3g and Fig. 3j) directions. Two peaks are observed below the Fermi level in both cases: the one closer to the Fermi level corresponds to the $\delta$ band while the other one with a higher binding energy corresponds to the $\varepsilon$ band. For the tiny electron-like $\delta$ band, the observed band is nearly flat in the superconducting state (Fig. 3f and Fig. 3l). It gives a superconducting gap size of 5.5\,$\pm$\,0.5\,meV for the $\delta$ band. Regarding the superconducting gap for the $\varepsilon$ band, first, this band is already observed in the normal state with the band position at the M point (k$_4$ in Fig. 3a and Fig. 3c) at $\sim$13\,meV (marked by black arrows in Fig. 3o and Fig. 3p) and the tip position of the M-shaped band (k$_6$ in Fig. 3a and Fig. 3c) at $\sim$11\,meV (marked by black arrows in Fig. 3q). The M-shaped $\varepsilon$ band (Figs. 3a-3c) does not cross the Fermi level in the normal state. In the superconducting state, although strong superconducting coherence peaks form on the $\varepsilon$ band, the overall energy position of the band shows only a slight shift compared to that in the normal state, as seen from the EDCs at both the M point (k$_4$) and the tip position of the M-shaped band (k$_6$) in Figs. 3o-3q. This indicates that the gap opening on the $\varepsilon$ band must be significantly smaller than the energy position of the M point (13\,meV) and the tip position of the M-shaped band (11\,meV). If there is any gap opening on the $\varepsilon$ band, the estimated gap size should be smaller than 6\,meV (estimated from the peak position of the EDCs at $k_4$ (Figs. 3o and 3p) and $k_6$ (Fig. 3q) in the normal state (13\,meV for $k_4$ and 11\,meV for $k_6$) and superconducting state (14\,meV for $k_4$ and 11\,meV for $k_6$)). We note that the superconducting gap we obtained for both the $\delta$ band ($\sim$5.5\,meV) and the $\varepsilon$ ($<$\,6\,meV) band near the M point in the optimally-doped (Ba$_{0.6}$K$_{0.4}$)Fe$_2$As$_2$ superconductor is significantly smaller than the values (10-12\,meV) reported in the previous studies\cite{XJZhou2008LZhao, HDing2008NLWang, MZHasan2008LWray, TTakahashi2009KNakayama, SShin2017TShimojima}. Since we can observe the $\delta$ and $\varepsilon$ bands simultaneously both in the normal state and in the superconducting state, we can make a proper assignment of the bands and keep track on their positions in both the normal and superconducting states. Therefore, the gap size $\sim$5.5\,meV we obtained for the $\delta$ band is reliable and represents the intrinsic superconducting gap structure near the M point. When the band structure in the normal state is not measured or only one band is observed near the M point in the superconducting state, the energy position of the $\varepsilon$ band may be mistakenly taken as the superconducting gap size, either at the M point or at the tip position of the M-shaped band.

In order to check on the possible folding of bands from M to $\Gamma$ in the superconducting state of the optimally-doped (Ba$_{0.6}$K$_{0.4}$)Fe$_2$As$_2$ superconductor\cite{SShin2017TShimojima}, we directly compare the band structures along the same directions crossing $\Gamma$ and M. Figs. 4a-4d show the comparison of the two bands measured along the $\Gamma$-M$_1$ direction while Figs. 4e-4h show the two bands measured along the $\Gamma$-X direction. For each band measurement, we have identified the observed bands and marked them in the second derivative images (Figs. 4c, 4d, 4g and 4h). Fig. 4i directly compares EDCs at two typical momentum points around the $\Gamma$ point ($\Gamma$k$_1$, $\Gamma$k$_2$ in Fig. 4a and $\Gamma$k$_3$, $\Gamma$k$_4$ in Fig. 4e) with those at the two equivalent momentum points around the M point (Mk$_1$, Mk$_2$ in Fig. 4b and Mk$_3$, Mk$_4$ in Fig. 4f). The features corresponding to various bands are marked in the EDCs. In the band structure measured along the $\Gamma$-M$_1$ direction crossing the $\Gamma$ point (Fig. 4a and Fig. 4c), in addition to the regular hole-like $\alpha$/$\beta$ and $\gamma$ bands, we can identify another two bands around the $\Gamma$ point, $\delta'$ band (dashed thin cyan line in Fig. 4c) and $\varepsilon'$ band (dashed thin red line in Fig. 4c). The $\varepsilon'$ band crosses the $\alpha$/$\beta$ band and extends to the region between the $\alpha$/$\beta$ band and the $\gamma$ band, as seen directly from the measured band structure (Fig. 4a and Fig. 4c) and the EDC at $\Gamma$k$_2$ in Fig. 4i. The shape and position of the $\delta'$ and $\varepsilon'$ bands around the $\Gamma$ point show a strong resemblance to the $\delta$ and M-shaped $\varepsilon$ bands around the M point in Fig. 4d. In the band structure measured along the $\Gamma$-X direction crossing the $\Gamma$ point (Fig. 4e and Fig. 4g), in addition to the hole-like $\alpha$/$\beta$ and $\gamma$ bands, we can also identify another two bands around the $\Gamma$ point, $\delta'$ band (dashed thin cyan line in Fig. 4g) and $\varepsilon'$ band (dashed thin red line in Fig. 4g).  The shape and position of the $\delta'$ and $\varepsilon'$ bands around the $\Gamma$ point also exhibit a strong similarity to the $\delta$ and $\varepsilon$ bands around the M point in Fig. 4h. Note that, when the band at Mk$_4$ is absent (Fig. 4f, Fig. 4h and the EDC at Mk$_4$ in Fig. 4i), it also becomes absent at $\Gamma$k$_4$ (Fig. 4e, Fig. 4g and the EDC at $\Gamma$k$_4$ in Fig. 4i). All these results indicate that there is a band folding from M to $\Gamma$ in the superconducting state of the optimally-doped (Ba$_{0.6}$K$_{0.4}$)Fe$_2$As$_2$ superconductor.

We find that the band folding alone from M to $\Gamma$ can not fully account for the flat band feature around the $\Gamma$ point. First, if the flat band near the $\Gamma$ point comes only from the band folding of the $\varepsilon$ band around the M point, one would expect that the band position of the flat band at $\Gamma$ is exactly the same as the position of the $\varepsilon$ band at the M point. But we find that these two positions do not exactly match; there is a feature in the EDCs at $\Gamma$ which has a higher binding energy than the EDC peak position at M, as seen in the EDCs at ($\Gamma$k$_1$,Mk$_1$) and ($\Gamma$k$_3$,Mk$_3$) in Fig. 4i. Second, as seen in Fig. 4c and Fig. 4g, around the $\Gamma$ point, there are additional features observed below the folded bands, $\varepsilon'$. Third, a careful examination of the EDCs around the $\Gamma$ point indicates that they are composed of two features, as marked by the black and red tips in Fig. 1c. All these observations indicate that the flat band around the $\Gamma$ point consists of two components, one is the folded band from the $\varepsilon$ band around M, the other is from the Bogoliubov back-bending of the $\alpha$/$\beta$ band as shown in Fig. 4c and Fig. 4g. There are three hole-like bands around the Brillouin zone center, $\alpha$, $\beta$ and $\gamma$, that exhibit distinct k$_z$ dependence\cite{DLFeng2011YZhang,HDing2011YMXu}. For the photon energy of 21.218\,eV, the corresponding k$_z$ is close to zero, and the $\alpha$ and $\beta$ bands are nearly degenerate\cite{DLFeng2011YZhang,HDing2011YMXu}. In the superconducting state, these two bands may have different superconducting gap size; the $\beta$ band shows a larger superconducting gap than the $\alpha$ band as seen in Fig. 4c and Fig. 4g, consistent with the previous result\cite{DLFeng2011YZhang}.

As we have shown above, there is band folding from M to $\Gamma$ in the superconducting state of the optimally-doped (Ba$_{0.6}$K$_{0.4}$)Fe$_2$As$_2$ superconductor. If such a band folding exists, it is natural to ask whether there is also band folding from $\Gamma$ to M, and furthermore, whether such a band folding also occurs in the normal state. Fig. 5 directly compares the band structures along the same directions crossing $\Gamma$ and M in the normal and superconducting states. Figs. 5a-5d show the comparison of the two bands measured along the $\Gamma$-M$_1$ direction in the normal and superconducting states while Figs. 5e-5h show the two bands measured along the $\Gamma$-X direction. In all these four cases, we can observe signatures of bands around M that are folded from the bands around $\Gamma$. The band folding of the $\gamma$ band around $\Gamma$ is particularly clear as the folded band around the M point can be clearly seen, as marked by the red arrows in Fig. 5b, Fig. 5d, Fig. 5f and Fig. 5h. These results have provided further evidence of the band folding between $\Gamma$ and M in both the normal state and superconducting state in the optimally-doped (Ba$_{0.6}$K$_{0.4}$)Fe$_2$As$_2$ superconductor.

We have shown that the flat band feature around $\Gamma$ consists of two components: the folded $\varepsilon$ band from M ($\varepsilon'$ in Fig. 4c and Fig. 4g) and the back-bending band of the $\beta$ band ($\beta$ band in Fig. 4c and Fig. 4g). In the measured band structures around the $\Gamma$ point using Helium lamp \emph{h}$\nu$=21.218\,eV (Fig. 1b, Fig. 4a, Fig. 4c, Fig. 4e and Fig. 4g), the $\beta$ back-bending component of the flat band is stronger than the folded component. On the other hand, in the measured band structures around the $\Gamma$ point using laser \emph{h}$\nu$=6.994\,eV (Figs. 2c-2f), the folded component of the flat band is dominant. Fig. 6a shows the dispersions of the back-bending component of the flat band obtained from the Helium lamp measurement (Fig. 4a) and laser measurement (Fig. 2d). For comparison, the dispersion of the $\varepsilon$ band from Helium lamp measurement (Fig. 4b) is also plotted in Fig. 6a. These dispersions, measured along the $\Gamma$-M$_1$ direction, are M-shaped and can span a wide momentum space crossing the $\alpha$/$\beta$ band. They agree with each other in both the shape and the positions of energy and momentum. This further confirms that this component of the flat band around $\Gamma$ originates from the folding of the $\varepsilon$ band around M.

Based on the above results, we can now establish a coherent picture to understand the observed electronic structure of the optimally-doped (Ba$_{0.6}$K$_{0.4}$)Fe$_2$As$_2$ superconductor both near the $\Gamma$ region and the M region. The Fermi surface measured by Helium lamp \emph{h}$\nu$=21.218\,eV (k$_z$ = 0) consists of two hole-like Fermi surface sheets around $\Gamma$ and a tiny electron-like pocket at M (Fig. 6b). When there is a band folding between the $\Gamma$ and M points, the Fermi surface sheets will be folded between the two points (Fig. 6c). Whether the folded Fermi surface sheets can be seen depends on the folding strength. In the normal state without band folding, the band structure measured along the $\Gamma$-M direction consists of three hole-like bands around $\Gamma$ ($\alpha$, $\beta$ and $\gamma$, with $\alpha$ and $\beta$ being degenerate in this case), as well as a tiny electron-like band ($\delta$) and an M-shaped $\varepsilon$ band (Fig. 6d). The $\varepsilon$ band is rather incoherent in the normal state. Upon entering the superconducting state, superconducting gap opens and superconducting coherence peak forms and Bogoliubov back-bending bands appear on all these bands (Fig. 6e). Around the $\Gamma$ point, the initially degenerate $\alpha$ and $\beta$ bands open superconducting gap with different sizes; the gap size of the $\beta$ band is larger than that of the $\alpha$ band. Around the M point, the electron-like $\delta$ band produces a section of a flat band below the Fermi level and the initially incoherent M-shaped $\varepsilon$ band becomes rather coherent in the superconducting state (Fig. 6e). When the band folding is considered in the superconducting state, the original hole-like bands around $\Gamma$ are folded to the M region while the original bands around the M point are folded to the $\Gamma$ region (Fig. 6f). In this way, the flat band feature observed around $\Gamma$ in the superconducting state can be understood as arising from the combination of the folded $\varepsilon$ band from M and the back-bending of the $\beta$ band. In the meantime, two coherent bands are observed simultaneously around the M point. The folded $\varepsilon$ band around $\Gamma$ can be more clearly observed in the superconducting state because the incoherent $\varepsilon$ band around M in the normal state becomes rather coherent in the superconducting state.

The clarification of the electronic structures in the optimally-doped (Ba$_{0.6}$K$_{0.4}$)Fe$_2$As$_2$ superconductor is critical to make correct band assignment and precise determination of the related superconducting gap. As we have now understood, the degenerate $\alpha$ and $\beta$ bands in the normal state open superconducting gaps of different sizes. The EDC at the Fermi momentum from the measured band (Fig. 7a) is shown in Fig. 7b which consists of two peaks. The extracted gap size from the symmetrized EDC (Fig. 7d) is 12\,meV for the $\beta$ band and 7\,meV for the $\alpha$ band. The superconducting gap of the $\gamma$ band obtained from Fig. 7c and Fig. 7e is 5\,meV. It has been found that the superconducting gap of the hole-like bands around the $\Gamma$ point is highly isotropic\cite{XJZhou2019JWhuang}. Combining the superconducting gap results measured around the M point (Fig. 3o and Fig. 3p), we establish an overall superconducting gap structure for the optimally-doped (Ba$_{0.6}$K$_{0.4}$)Fe$_2$As$_2$ superconductor shown in Fig. 7f. The superconducting gap structure around the M point is distinct from all the previous measurements\cite{XJZhou2008LZhao, HDing2008NLWang, MZHasan2008LWray, TTakahashi2009KNakayama, SShin2017TShimojima}.

Our results prove that there is a band folding effect between the Brillouin zone center and the corners with a wave vector of (0,0)-($\pm\pi$,$\pm\pi$) in both the normal state and superconducting state of the optimally-doped (Ba$_{0.6}$K$_{0.4}$)Fe$_2$As$_2$ superconductor. The band folding from the zone corner to the zone center in the superconducting state was also reported before, which is attributed to some unknown electronic instability\cite{SShin2017TShimojima}. In the parent and underdoped BaFe$_2$As$_2$, there is a spin-density wave (SDW) order which plays an important role in determining the electronic structure\cite{XHChen2009HChen}. ARPES experiments have clearly shown a folding effect between the $\Gamma$ and M due to the SDW order \cite{DLFeng2009LXYang, XJZhou2009GDLiu, ZXShen2009MYi, AKaminski2010TKondo}. However, in the optimally-doped (Ba$_{0.6}$K$_{0.4}$)Fe$_2$As$_2$ superconductor, the SDW order is completely suppressed and can not explain the band structure folding between $\Gamma$ and M we observed. Recently, scanning tunneling spectroscopy revealed the presence of regions with a $\sqrt{2} \times \sqrt{2}$ buckling reconstruction in the optimally-doped (Ba$_{0.6}$K$_{0.4}$)Fe$_2$As$_2$ superconductor\cite{SHPan2019ALi}. Whether the band folding we observed can be due to this reconstruction or some other origins asks for further investigations.

In summary, we have carried out high-resolution ARPES measurements on the optimally-doped (Ba$_{0.6}$K$_{0.4}$)Fe$_2$As$_2$ to resolve the detailed electronic structure and superconducting gap structure. By using both Helium lamp and laser light sources, we observe a clear "flat band" feature around the $\Gamma$ point in the superconducting state. Our results indicate that this flat band originates from the combined effect of both the superconductivity-induced band back-bending of the $\beta$ band and the folding of the $\varepsilon$ band from M to $\Gamma$. Direct evidence of the band folding between M and $\Gamma$ is observed in both the normal state and superconducting state. Our resolution of the origin of the flat band around the $\Gamma$ point makes it possible to assign the three hole-like bands around $\Gamma$ and determine their superconducting gap properly. Near the M point, we observe both a tiny electron-like band $\delta$ and an M-shaped $\varepsilon$ band simultaneously in normal and superconducting states. This makes it possible to correctly determine the superconducting gap around the M point. The obtained superconducting gap for the bands around the M point ($\sim$5.5\,meV) is significantly different from the previous measurements. Our results resolve a number of prominent controversies concerning the electronic structure and superconducting gap structure in the prototypical iron-based superconductor, the optimally-doped (Ba$_{0.6}$K$_{0.4}$)Fe$_2$As$_2$. They provide key information in examining and establishing theories in understanding superconductivity mechanism in iron-based superconductors.

\vspace{3mm}


\vspace{3mm}

\noindent {\bf Acknowledgement}\\
We are thankful for financial support from the National Key Research and Development Program of China (Grants No. 2016YFA0300300, 2017YFA0302900, 2018YFA0704200 and 2019YFA0308000), the National Natural Science Foundation of China (Grants No. 11888101, 11922414, 11874405, 62022089), the Strategic Priority Research Program (B) of the Chinese Academy of Sciences (Grants No. XDB25000000, XDB33000000), the Youth Innovation Promotion Association of CAS (Grants No. 2017013, 2019007), and the Research Program of Beijing Academy of Quantum Information Sciences (Grant No. Y18G06).


\vspace{3mm}

\noindent {\bf Author Contributions}\\
 X.J.Z., L.Z., Y.Q.C. and J.W.H. proposed and designed the research. J.W.H. contributed in sample growth. Y.Q.C., J.W.H., T.M.M., D.S.W., Q.G., C.L., Y.X., J.J.J, Q.Y.W., Y.H., G.D.L., F.F.Z., S.J.Z., F.Y., Z.M.W., Q.J.P., Z.Y.X., L.Z. and X.J.Z. contributed to the development and maintenance of the ARPES systems and related software development. Y.Q.C. carried out the ARPES experiment with J.W.H.. Y.Q.C., L.Z. and X.J.Z. analyzed the data. Y.Q.C., L.Z. and X.J.Z. wrote the paper. All authors participated in discussion and comment on the paper.\\

\vspace{3mm}

\noindent{\bf Additional information}\\
Supplementary information is available in the online version of the paper.
Correspondence and requests for materials should be addressed to L. Zhao, and X. J. Zhou.

\newpage

\begin{figure*}[tbp]
\begin{center}
\includegraphics[width=1.0\columnwidth,angle=0]{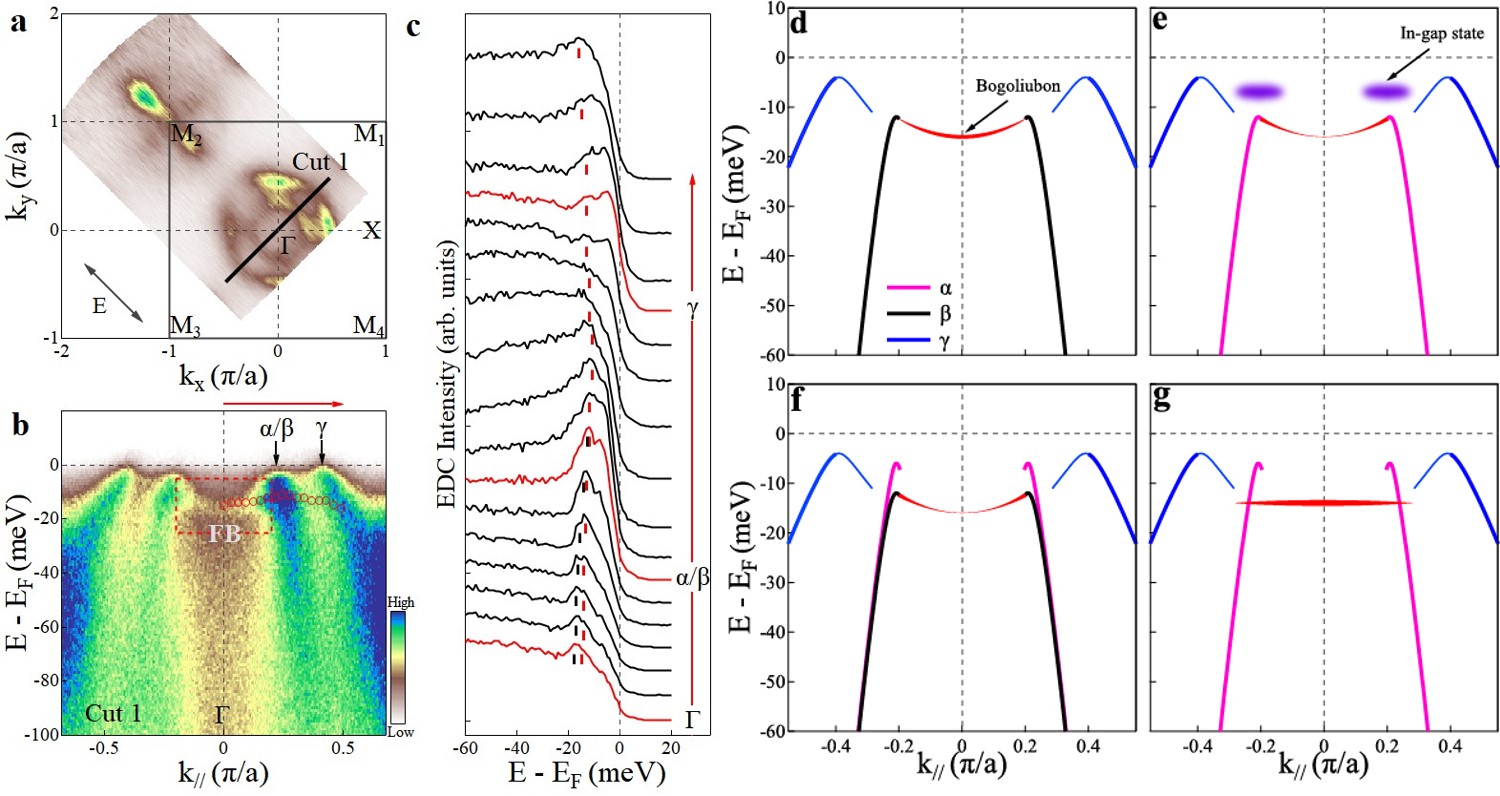}
\end{center}

\caption{{\bf Flat band (FB) observed near the $\Gamma$ point in the superconducting state of the optimally-doped (Ba$_{0.6}$K$_{0.4}$)Fe$_2$As$_2$ superconductor in Helium lamp ARPES measurement and its possible origins.} (a) Fermi surface mapping measured at 18\,K with \emph{h}$\nu$=21.218\,eV Helium lamp. It is obtained by integrating the spectral weight within a $\pm$ 1 meV energy window relative to the Fermi level E$_F$. The major polarization vector (E) is marked at the bottom-left corner. The first Brillouin zone is represented by a black square. For convenience, the equivalent four zone corners are named as $M_1$, $M_2$, $M_3$ and $M_4$. (b) Band structure measured along the $\Gamma$-$M_1$ direction crossing the $\Gamma$ point; the location of the momentum Cut 1 is marked in (a) by black line. There is a "flat band" formed near the $\Gamma$ point, as marked in the red rectangular frame. (c) Photoemission spectra (EDCs) from the band structure in (b). Red tips mark the peak position of the flat band that is also plotted in (b) by red circles. The peak position of the Bogoliubov back-bending band of the inner hole-like band is marked by black tips. (d-g) shows schematic of four possible scenarios that are proposed to understand the origin of the flat band. Three hole-like bands and the flat band are marked as pink, black, blue and red lines, respectively.  (d) The flat band is from the Bogoliubons of the $\beta$ band in the superconducting state. (e) Formation of the "in-gap" state and the flat band is from the Bogoliubov back-bending band of the $\alpha$ band\cite{HDing2014PZhang}. (f) The flat band is from the Bogoliubov back-bending band of the $\beta$ band in the superconducting state\cite{DLFeng2010YZhang}. In this case, $\alpha$ and $\beta$ bands are degenerate in the normal state but have different superconducting gaps in the superconducting state. (g) The flat band is from the band folding of the structure near M ($\pi$,$\pi$) point\cite{SShin2017TShimojima}.
}
\end{figure*}

\begin{figure*}[tbp]
\begin{center}
\includegraphics[width=1.0\columnwidth,angle=0]{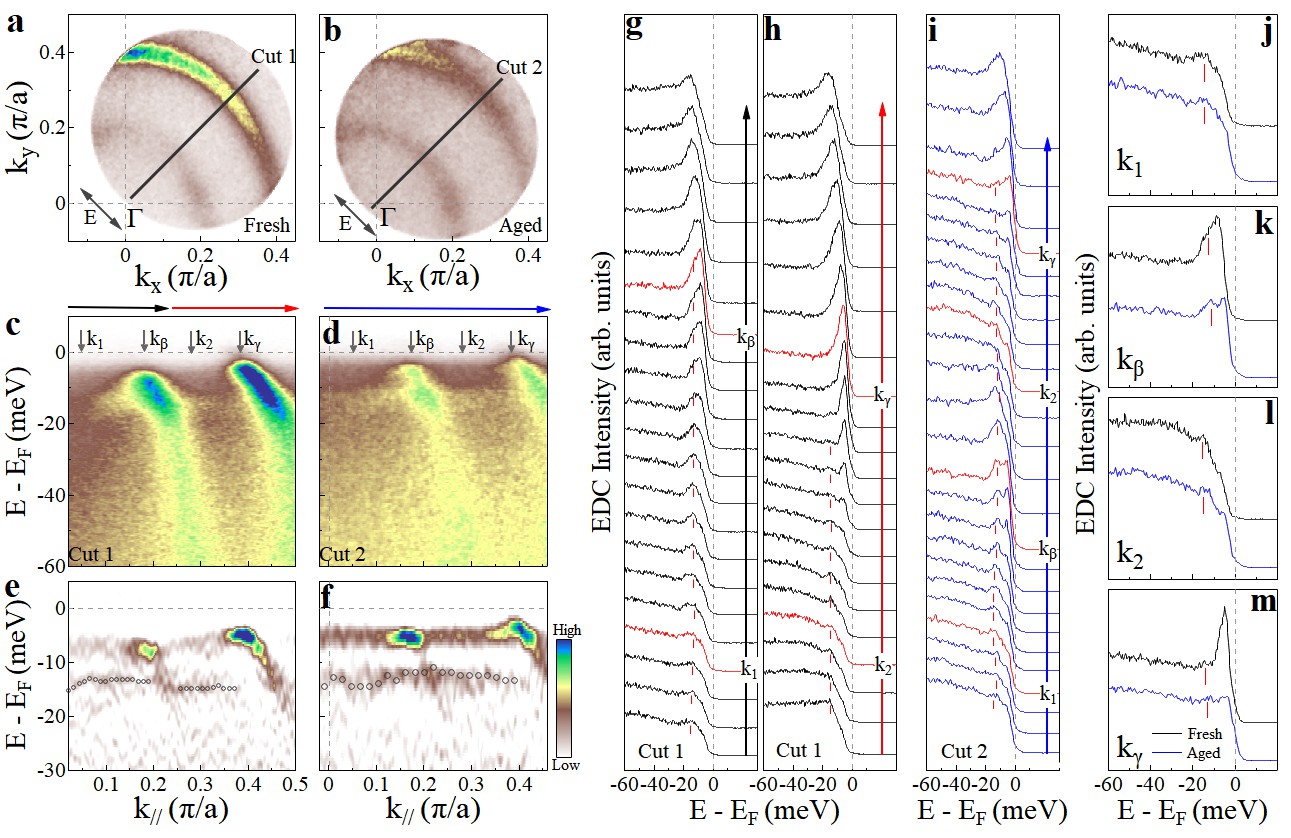}
\end{center}

\caption{{\bf Flat band observed near the $\Gamma$ point in fresh and aged (Ba$_{0.6}$K$_{0.4}$)Fe$_2$As$_2$ in the superconducting state in laser ARPES measurement.} (a-b) Fermi surface mapping of fresh (a) and aged (b) (Ba$_{0.6}$K$_{0.4}$)Fe$_2$As$_2$ at 13\,K with \emph{h}$\nu$=6.994\,eV laser. They are obtained by integrating the spectral weight within a $\pm$ 5 meV energy window relative to the Fermi level E$_F$. The polarization vector (E) is marked at the bottom-left corner. The data for the fresh sample (a) is obtained right after the cleaving. After that, the sample was heated up to high temperature up to 70\,K and cooled down to 13\,K again. The data for the aged sample (b) is then measured more than 24 hours after cleaving. (c-d) Band structure measured along the Cut 1 (c) and Cut 2 (d), respectively. The location of Cut 1 and Cut 2 is shown in (a) and (b) by black lines and they are along the same $\Gamma$-$M_1$ direction in the momentum space. (e-f) The second derivative images of (c) and (d) with respect to the energy. (g-h) Photoemission spectra (EDCs) from the band structure in (c). Red tips mark the peak position of the flat band that is also plotted in (e) by black circles. (i) EDCs from the band structure in (d). Red tips mark the peak position of the flat band that is also plotted in (f) by black circles. (j-m) Comparison of EDCs between the fresh and aged samples at four typical momenta: k$_1$ (j), k$_\beta$ (k), k$_2$ (l), k$_\gamma$ (m). The location of the four momenta is marked by arrows in (c) and (d).
}
\end{figure*}

\begin{figure*}[tbp]
\begin{center}
\includegraphics[width=1.0\columnwidth,angle=0]{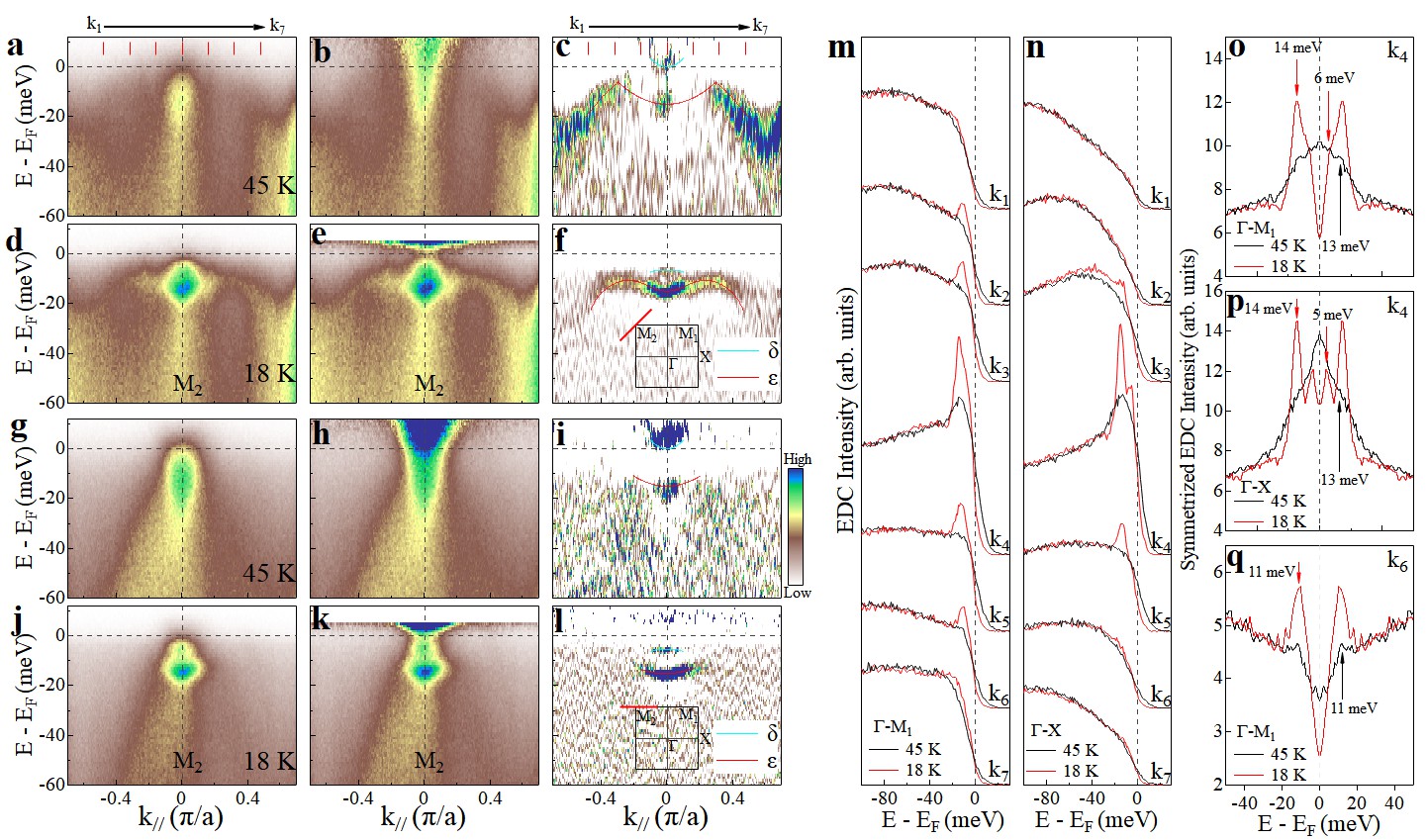}
\end{center}

\caption{{\bf Detailed band structure of (Ba$_{0.6}$K$_{0.4}$)Fe$_2$As$_2$ near the M point in both normal and superconducting states measured by \emph{h}$\nu$=21.218\,eV Helium lamp.} (a) Band structure measured along $\Gamma$-$M_1$ direction crossing $M_2$ point at 45\,K in the normal state. The location of the momentum cut is shown by the red line in the inset of (f). (b) Band structure of (a) divided by the corresponding Fermi distribution function. (c) Second derivative image of (b) with respect to the energy. (d-f) Same as (a-c) but measured at 18\,K in the superconducting state. (g-i) Same as (a-c) but measured along $\Gamma$-X direction crossing $M_2$ point at 45\,K in the normal state. The location of the momentum cut is shown by the red line in the inset of (l). (j-l) Same as (g-i) but measured at 18\,K in the superconducting state. The observed two bands, $\delta$ and $\varepsilon$, are marked by cyan and red lines, respectively, in the second derivative images of (c), (f), (i) and (l). (m) EDCs from the band structures (a) and (d) at several momenta measured at 45\,K (black lines) and 18\,K (red lines). The location of the momentum points is marked by red tips in (a). (n) Same as (m) but from the band structures (g) and (j). (o) Symmetrized EDCs at 45\,K (black line) and 18\,K (red line) from the original EDCs for the momentum $k_4$ in (m). (p) Symmetrized EDCs at 45\,K (black line) and 18\,K (red line) from the original EDCs for the momentum $k_4$ in (n). (q) Symmetrized EDCs at 45\,K (black line) and 18\,K (red line) from the original EDCs for the momentum $k_6$ in (m).
}
\end{figure*}

\begin{figure*}[tbp]
\begin{center}
\includegraphics[width=1.0\columnwidth,angle=0]{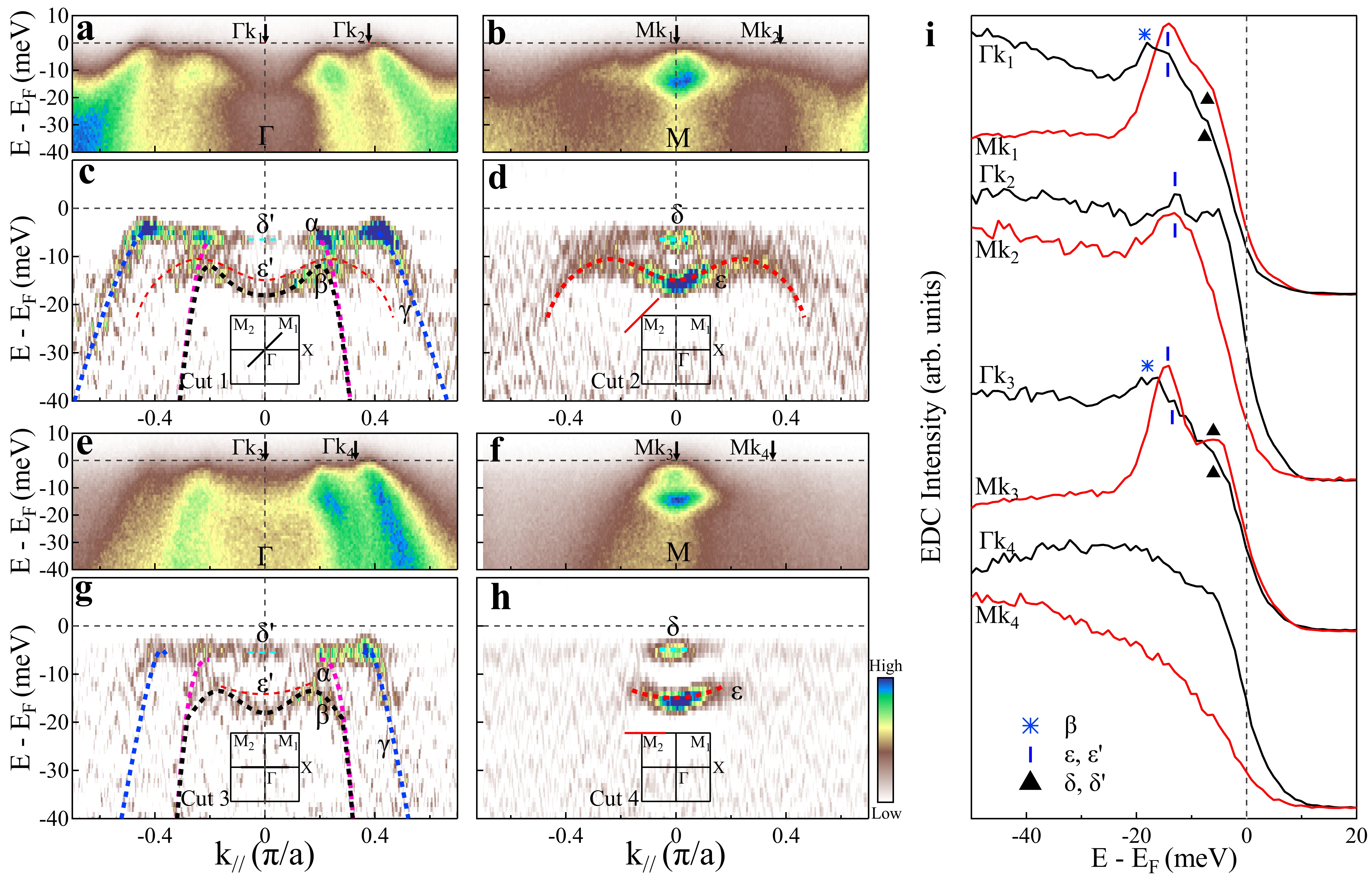}
\end{center}

\caption{{\bf Detailed comparison of band structures of (Ba$_{0.6}$K$_{0.4}$)Fe$_2$As$_2$ between $\Gamma$ and M measured at 18\,K in the superconducting state.} (a-b) Band structures measured along the same $\Gamma$-$M_1$ direction crossing the $\Gamma$ point (a) and $M_2$ point (b). (c-d) Corresponding second derivative images of (a) and (b) with respect to the energy. The location of the momentum cuts is marked in the inset of (c) and (d). (e-h) Same as (a-d) but measured along the $\Gamma$-X direction. The location of the momentum cuts is marked in the inset of (g) and (h). The observed $\alpha$, $\beta$ and $\gamma$ bands at the $\Gamma$ point are marked by pink, black and blue thick dashed lines, respectively, while the bands $\delta$ and $\varepsilon$ at the M point are marked by cyan and red thick dashed lines. The $\delta'$ and $\varepsilon'$ bands marked by cyan and red thin dashed lines at the $\Gamma$ point in (c) and (g) represent the folded bands from the $\delta$ and $\varepsilon$ bands at the M point. (i) Comparison of EDCs at the corresponding momenta around $\Gamma$ and M. The location of the four pairs of momenta, (${\Gamma}k_1$, $Mk_1$), (${\Gamma}k_2$, $Mk_2$), (${\Gamma}k_3$, $Mk_3$) and (${\Gamma}k_4$, $Mk_4$), is marked in (a), (b), (e) and (f). The three features, labelled as star, tip and triangle, correspond to $\beta$, $\varepsilon$ and $\delta$ bands, respectively.
}

\end{figure*}

\begin{figure*}[tbp]
\begin{center}
\includegraphics[width=1.0\columnwidth,angle=0]{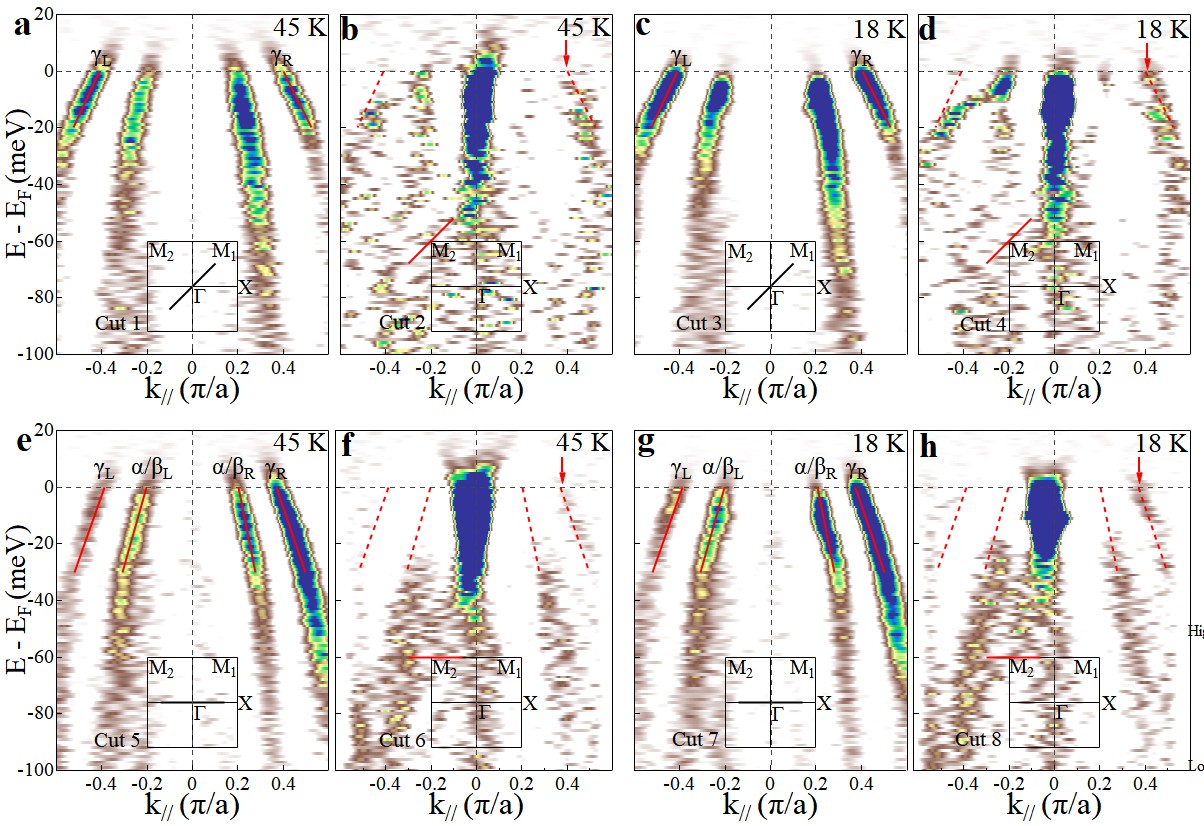}
\end{center}

\caption{{\bf Signatures of band folding from $\Gamma$ to the M in (Ba$_{0.6}$K$_{0.4}$)Fe$_2$As$_2$ both in the normal state and superconducting state.} (a-b) Band structures measured along the same $\Gamma$-$M_1$ direction crossing the $\Gamma$ point (a) and $M_2$ point (b) at 45\,K in the normal state. These are second derivative images with respect to momentum. (c-d) Same as (a-b) but measured at 18\,K in the superconducting state. (e-h) Same as (a-d) but measured along the $\Gamma$-X direction. The location of the momentum cuts is marked in the inset. The dashed red lines in the bands near the M point represent the position of the bands folded from the $\Gamma$ point (solid red lines).
}
\end{figure*}

\begin{figure*}[tbp]
\begin{center}
\includegraphics[width=1.0\columnwidth,angle=0]{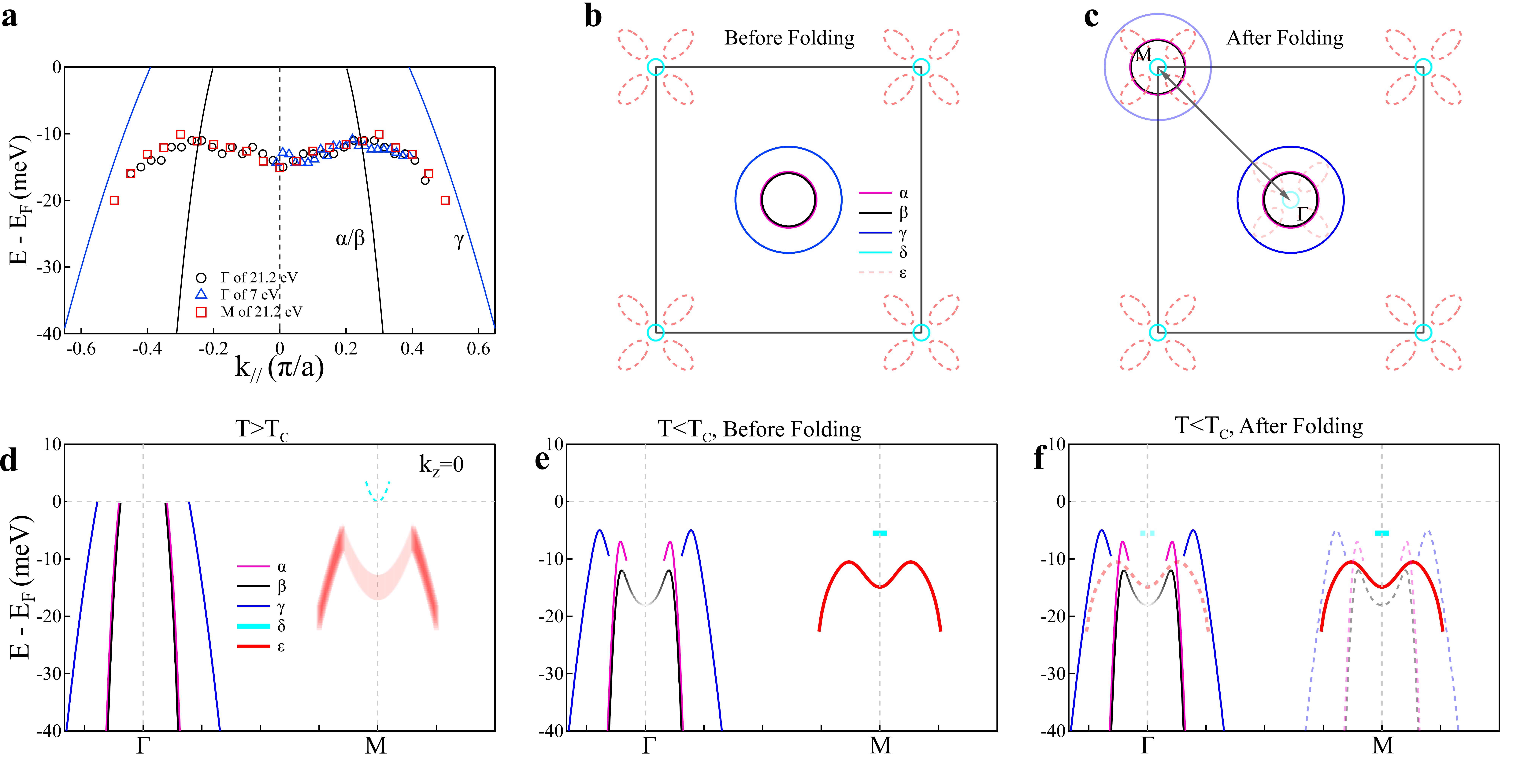}
\end{center}

\caption{{\bf Formation of the flat band in (Ba$_{0.6}$K$_{0.4}$)Fe$_2$As$_2$ at the $\Gamma$ point in the superconducting state from the Bogoliubov back-bending of the $\beta$ band and the folding of the $\varepsilon$ band at M to $\Gamma$.} (a) Quantitative comparison of the flat band at the $\Gamma$ point with the $\varepsilon$ band at the M point. The black circles represent the flat band dispersion at $\Gamma$ extracted from Helium lamp measurement (\emph{h}$\nu$=21.218\,eV) in Fig. 4a, and the blue triangles represent the flat band dispersion at the $\Gamma$ point from laser measurement (\emph{h}$\nu$=6.994\,eV) in Fig. 2d. The red squares represent the $\varepsilon$ band dispersion at the M point from Helium lamp measurement (\emph{h}$\nu$=21.218\,eV) in Fig. 3e. (b-c) Schematic Fermi surface of (Ba$_{0.6}$K$_{0.4}$)Fe$_2$As$_2$ at k$_z$ = 0 before (b) and after (c) $\Gamma$-M band folding. Since $\alpha$ and $\beta$ bands are degenerate, there are two hole-like pockets around the $\Gamma$ point and a small electron pocket around the M point. Four strong spots are illustrated by red dashed lines but the corresponding M-shaped bands lie below the Fermi level. The folded Fermi pockets are marked by solid thin lines in (c). (d) Schematic band structure of (Ba$_{0.6}$K$_{0.4}$)Fe$_2$As$_2$ in the normal state above T$_c$ at k$_z$ = 0. (e) Corresponding band structure in the superconducting state below T$_c$. (f) Band structure in the superconducting state after $\Gamma$-M folding. The folded bands are shown by the dashed lines.
   }

\end{figure*}

\begin{figure*}[tbp]
\begin{center}
\includegraphics[width=1.0\columnwidth,angle=0]{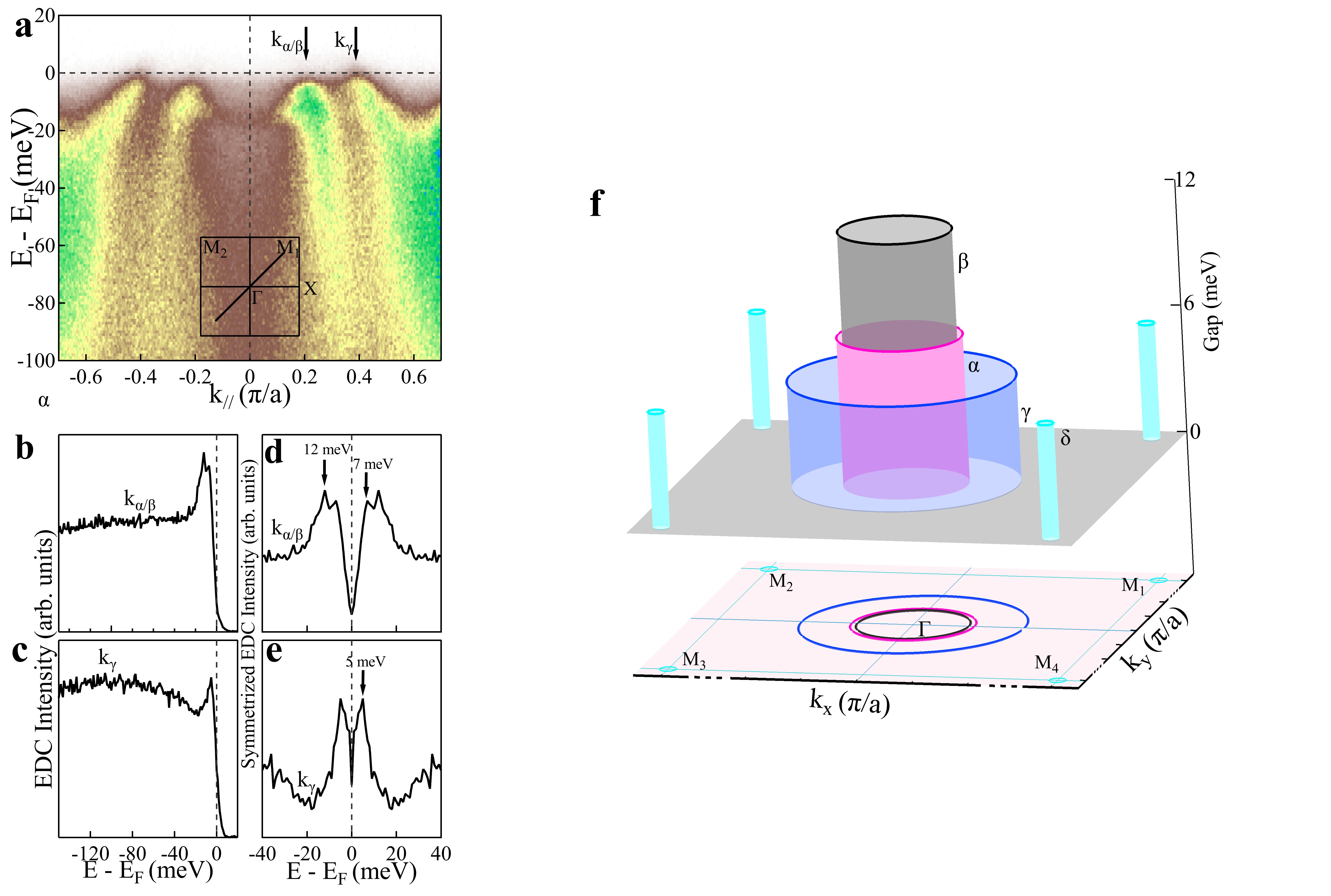}
\end{center}

\caption{{\bf The overall superconducting gap structure of the optimally-doped (Ba$_{0.6}$K$_{0.4}$)Fe$_2$As$_2$ superconductor.} (a) Band structure measured at 18\,K by \emph{h}$\nu$=21.218\,eV Helium lamp along the $\Gamma$-$M_1$ direction crossing the $\Gamma$ point. The location of the momentum cut is marked by a black line in the inset. (b-c) EDCs at $k_{\alpha/\beta}$ (b) and at $k_{\gamma}$ (c). (d-e) Corresponding symmetrized EDCs at $k_{\alpha,\beta}$ (d) and at $k_{\gamma}$ (e). The location of the $k_{\alpha/\beta}$ and $k_{\gamma}$ is marked by arrows in (a). (f) Three-dimensional plot of the superconducting gap on the three hole-like pockets around $\Gamma$ and the tiny electron-like pockets around M. The corresponding Fermi surface is shown at the bottom.
}
\end{figure*}

\end{document}